\documentclass[12pt]{article}
\usepackage{epsfig}

\newcommand{\beqn}{\begin{eqnarray}}
\newcommand{\eeqn}{\end{eqnarray}}
\newcommand{\eq}[1]{(\ref{#1})}
\newcommand{\dual}[1]{{}^{*}{#1}}
\newcommand{\dd}{\mathrm{d}}

\newcommand{\itep}
{~\vspace{-1.5cm}
\begin{flushright}
{\large ITEP-LAT/2004-08}\\
\end{flushright}
\vspace{1.0cm}}

\begin{document}
\baselineskip=14pt
\begin{center}

\itep

{\Large\bf On projection (in)dependence of monopole condensate in lattice SU(2)
gauge theory}

\vskip 1.0cm {\bf V.A.~Belavin, M.N.~Chernodub and
M.I.~Polikarpov}
\vskip 4mm
{\it ITEP, B.~Cheremushkinskaya 25, Moscow, 117259, Russia}

\end{center}

\begin{abstract}
We study the temperature dependence of the monopole condensate in different
Abelian projections of the $SU(2)$ lattice gauge theory. Using the
Fr\"{o}hlich-Marchetti monopole creation operator we show numerically that the
monopole condensate depends on the choice of the Abelian projection. Contrary
to the claims in the current literature we observe that in the Abelian Polyakov
gauge and in the field strength gauge the monopole condensate does not vanish
at the critical temperature and thus is not an order parameter.
\end{abstract}

\section{Introduction}

The confinement of color in QCD is one of the most interesting issues in the
modern quantum field theory. Numerical simulations of non--Abelian gauge
theories on the lattice~\cite{Bali:1994de} show that the confinement of quarks
happens due to a formation of the chromoelectric strings spanned between quarks
and anti-quarks. Although an analytical derivation of the color confinement is
not available yet, there are various effective models which describe the
emergence of the string. According to the dual superconductor
model~\cite{DualSuperconductor}, the vacuum of a non--Abelian gauge model may
be regarded as a medium of condensed Abelian monopoles. The monopole condensate
squeezes the chromoelectric flux (coming from the quarks) into a flux tube due
to the dual Meissner effect. This flux tube is an analogue of the Abrikosov
vortex in an ordinary superconductor.

The basic element of the dual superconductor picture is the Abelian monopole. This object
does not exist in QCD on the classical level, but is can be identified with
a particular configuration of the gluon fields with the help of the so--called Abelian
projection~\cite{AbelianProjections}. The Abelian projection uses a partial gauge fixing
of the SU(N) gauge symmetry up to an Abelian subgroup. The compactness of the residual
Abelian subgroup guarantees the existence of the Abelian monopoles in the Abelian projection.

Many numerical simulations show that the Abelian degrees of freedom in an
Abelian projection are responsible for the confinement of quarks (for a review,
see, $e.g.$, Refs.~\cite{Reviews}). One of the striking features of the Abelian
projection is the effect of the Abelian dominance~\cite{AbelianDominance}: the
Abelian gauge fields provide a dominant contribution to the tension of the
confining string. Moreover, the internal structure of the string energy, such
as energy profile and the field distribution are very well described by the
dual superconductor model~\cite{Bali:1994de}.

Since qualitative features of the confinement mechanism in QCD and in the pure
$SU(2)$ gauge theory are expected to be the same, below we restrict ourselves
to the simplest case of the $SU(2)$ gluodynamics. Most of the results
supporting the dual superconductor scenario were obtained in the so called
Maximal Abelian (MA) projection~\cite{kronfeld}. According to numerical
simulations~\cite{MonopoleCondensation1,MonopoleCondensation2} the monopole
condensate in the MA gauge is formed in the low temperature (confinement) phase
and the condensate disappears in the high temperature (deconfinement) phase in
the perfect agreement with the expectations coming from the dual superconductor
scenario.

Besides the MA projection there are Abelian projections which are defined by a
diagonalization of certain adjoint operators $X[U]$ with respect to the $SU(2)$
gauge transformations~\cite{AbelianProjections}. After the Abelian projection
is fixed, the matrix $X[U]$ becomes diagonal and the theory possesses the
(residual) $U(1)$ gauge symmetry. The most popular examples of such gauges are
the Abelian Polyakov (AP) gauge and the Abelian field strength gauge ($F_{12}$
gauge). These gauges correspond, respectively, to the diagonalization of the
Polyakov line and the $U_{12}$ plaquette variable.

One may ask whether a dual superconductor nature of non--Abelian vacuum is
realized in all Abelian gauges. Needless to say that this question is important
for understanding of the properties of the QCD vacuum. Indeed, it seems natural
that the color confinement must be described by a projection--independent model
since the confinement is a gauge invariant phenomenon. On the other hand one
can consider the Abelian projection as a gauge--dependent tool to associate the
confining  gluon configurations with the Abelian monopoles. This tool may work
well in one gauge and may not work in another gauge.

There are conflicting reports of the projection independence of the dual
superconductor mechanism of the color confinement\footnote{A brief review of
the current literature on this subject can be found in
Ref.~\cite{ref:Polyakov:Gauge}.}. The Abelian and Monopole dominance were
observed in more than one gauge~\cite{AbelianDominance,ref:IR:monopoles}. The
London penetration length measured in the MA projection is the same as the one
obtained without gauge fixing~\cite{Cea:1994ed}. The monopole condensation
studied with the help of a monopole creation operator was observed not only in
the MA gauge of $SU(2)$ gauge
theory~\cite{MonopoleCondensation1,MonopoleCondensation2} but also in other
gauges~\cite{DiGiacomo}.

On the other hand there are indications that the monopole
dynamics is affected by the choice of the Abelian projection.
In the MA gauge the monopole trajectories percolate only in the
confinement phase contrary to the case without any gauge
fixing, in which the monopoles are percolating in any phase~\cite{Misha}.
The chromoelectric string in different Abelian projections
looks differently: the correlation length (the inverse monopole
mass) extracted from the string profile in the AP gauge is
consistent with zero contrary to the MA
gauge~\cite{Bernstein:1996vr}. The chiral condensate is
dominated by the contributions of the Abelian monopoles in the
MA gauge~\cite{Woloshyn:1994rv,Lee:1995ac} contrary to
$F_{12}$~\cite{Woloshyn:1994rv} and AP~\cite{Lee:1995ac}
gauges.

One can show analytically that in the AP gauge the dual superconductor
mechanism can not be realized~\cite{ref:Polyakov:Gauge}. The reason is very
simple: the monopoles in the continuum limit in this gauge are static and they
can not contribute to the potential between static heavy quarks. On the other
hand it was shown in Ref.~\cite{ref:Polyakov:Gauge} that the absence of the
dual superconductor description in AP gauge does not contradict to the
statement of Ref.~\cite{DiGiacomo} that the monopole condensation is realized
in any gauge.

In this paper we are studying the effective potential for the monopole field
using the Fr\"{o}hlich-Marchetti~\cite{FrMa87} monopole creation operator. We
evaluate the potential in AP and $F_{12}$ gauges and compare the results with
the monopole potential obtained in the MA gauge~\cite{MonopoleCondensation1}.
The value of the monopole condensate is defined by the minimum of the effective
potential. We describe the monopole creation operator in
Section~\ref{sec:creation} and present numerical results in
Section~\ref{sec:numerics}. Our conclusions are summarized in the last Section.

\section{Abelian monopole creation operator in SU(2) model}
\label{sec:creation}

We study the $SU(2)$ gauge theory with the standard
Wilson action, $S[U]  = - 1/2 \sum_P \mbox{Tr}\, U_P$, where the sum
goes over the plaquettes $P$ and $U_P$ is the $SU(2)$ plaquette variable
composed of the link fields $U_l$, $U_P = U_1 U_2 U_3^\dagger
U_4^\dagger$. The link field is parameterized in the standard way:
\beqn
U_{x\mu} =\left(
\begin{array}{ll}
\cos \phi_{x\mu}\, e^{i\theta_{x\mu}} & \sin \phi_{x\mu}\, e^{i\chi_{x\mu}} \\
- \sin \phi_{x\mu}\, e^{- i\chi_{x\mu}} & \cos \phi_{x\mu}\, e^{i\theta_{x\mu}} \\
\end{array}
\right)\,,
\eeqn
$0 \le \phi \le \pi/2$ and $ -\pi < \theta,\chi \le \pi$.

In Abelian projection the residual gauge transformation matrices have the
diagonal form $\Omega^{\mathrm{Abel}}(\omega) = {\mathrm{diag} (e^{i
\omega},e^{- i \omega})}$, where $\omega$ is an arbitrary scalar function.
Under these transformations the diagonal field $\theta$ transforms as an
Abelian gauge field, $\theta_{x\mu} \to \theta_{x\mu} +\omega_x
-\omega_{x+\hat{\mu}}$, the off-diagonal field $\chi$ changes as a double
charged matter field, $\chi_{x\mu} \to \chi_{x\mu} +\omega_x +
\omega_{x+\hat{\mu}}$, the field $\phi$ remains intact. The $SU(2)$ plaquette
action contains~\cite{ref:minopoles} various interactions between these fields
as well as the action for the Abelian gauge field $\theta$: \beqn S[U] = -
\sum_P \beta_P[\phi] \cos \theta_P + \dots\,. \label{S} \eeqn Here $\theta_P  =
\theta_1 + \theta_2 - \theta_3 - \theta_4$ is a lattice analogue of the Abelian
field strength tensor and $\beta_P[\phi]$ is an effective coupling constant
dependent of the fields $\phi$, Ref.~\cite{ref:minopoles}.

Following Ref.~\cite{MonopoleCondensation1} we apply the monopole creation
operator of Fr\"{o}hlich-Marchetti~\cite{FrMa87} to the Abelian part of the
non--Abelian action~\eq{S}. Effectively, this operator shifts the Abelian
plaquette variable $\theta_P$ as follows\footnote{Note that in this paper we
are using the "old" definition~\cite{FrMa87} of the monopole creation operator.
The "new" definition~\cite{ref:FM:new} takes into account charged matter fields
but it is very involved from the point of view of numerical calculations.
Moreover, results of Ref.~\cite{MonopoleCondensation:Belavin:AHM} clearly show
that there is no qualitative difference between the old and the new
definitions.}:
\beqn
\Phi_{mon}(x) = \exp \left\{ \sum_P \beta_P[\phi] \biggl[
- \cos \theta_P + \cos(\theta_P + W_P(x)) \biggr] \right\}\,, \label{Ux2}
\eeqn
where $W_P = 2 \pi \delta\Delta^{-1}(D_x-\omega_x)$, $\omega_x$ is a Dirac
string attached to the monopole and the Dirac cloud $D_x$ satisfies the
equation $\delta \dual D_x = \dual \delta_x$.

We have used the differential form notations on the lattice described in detail
in the second paper in Ref.~\cite{Reviews}. $\delta$ ($\dd$) is the backward
(forward) derivative on the lattice which decreases (increases) by one the rank
of the form on which it acts. The rank of the form is determined by a
dimensionality of the lattice cell on which this form is defined. For example,
a scalar function is a 0-form, the vector function is a 1-form {\it etc}. If
$A$ is a lattice vector, then $\delta A$ is a scalar (a lattice analogue of the
divergence, $\partial_\mu A_\mu$) while $\dd A$ is an antisymmetric tensor (a
lattice analogue of the field strength, $\partial_{[\mu,} A_{\nu]}$). $\delta$
and $\dd$ operators are nilponent, {\it i.e.}, $\delta^2=0=\dd^2$. The lattice
Laplacian is $\Delta = \delta \dd + \dd \delta$, and $\Delta^{-1}$ is the
inverse Laplacian. The lattice Kronecker symbol is denoted as $\delta_x$: it is
a scalar function which is equal to unity at the site $x$ and zero otherwise.
The *-operator relates the forms on the dual and original lattice. For example,
if on the four-dimensional lattice $B$ is a scalar function (0-form) on the
original lattice, then $\dual B$ is a 4-form on the dual lattice.

The operator~\eq{Ux2} is clearly gauge invariant with respect to the $U(1)$
gauge transformations. One can also perform a formal duality transformation
with respect to the quantum average of the operator~\eq{Ux2} and show that in
the dual model -- which has form of the Abelian Higgs model -- this operator is
invariant under the (dual) gauge transformations~\cite{MonopoleCondensation1}.
Moreover, one can represent the partition function as a sum over closed
monopole trajectories. In this representation the quantum average of the
monopole creation operator $\Phi_{mon}(x)$, Eq.~\eq{Ux2}, is given by a sum
over all closed monopole trajectories plus one open trajectory which begins at
the point $x$, Refs.~\cite{FrMa87}. Thus, this operator creates a monopole at
the point $x$.

To get the monopole condensate we have to study the effective constraint
potential for the monopole creation operator $\Phi_{mon}$, \beqn V_{effc}(\Phi)
= - \ln (\langle \delta (\Phi - \frac{1}{V}\sum_x \Phi_{mon}(x))\rangle)\,.
\label{eq:Veff} \eeqn This potential selects the zero--momentum component of
the creation operator. The minimum of this potential corresponds to the
monopole condensate. However, a numerical calculation of the potential
$V_{effc}(\Phi)$ is time consuming, and in this paper we present results for
the probability distribution \beqn V(\Phi) = -\ln(\langle\delta (\Phi -
\Phi_{mon}(x))\rangle)\,, \label{eq:V} \eeqn which has a meaning very similar
to \eq{eq:Veff}.

We perform our study in the Abelian Polyakov gauge and in the Abelian field
strength gauge which are defined as we already discussed by the diagonalization
of the (untraced) Polyakov loop, $P_x[U]$, and of the $U_{x,12}$ plaquette,
\beqn
P_{x}[U] = U_{x,4} U_{x +\hat 4,4} \dots U_{x -\hat4,4}\,, \quad U_{x,12}
= U_{x,1} U_{x+\hat 1,2} U^\dagger_{x+\hat 2,1} U^\dagger_{x,2}\,,
\eeqn
with respect to gauge transformations, $U_{x,\mu} \to \Omega_x U_{x,\mu}
\Omega^\dagger_{x+\hat\mu}$.

We compare the potential in these gauges with the monopole potential obtained
in the MA gauge in Ref.~\cite{MonopoleCondensation1}. The MA gauge is defined
by the maximization of the lattice functional
\beqn
\max_{\Omega} R_{MA}[U^{\Omega}]\,, \qquad R_{MA}[U] = \sum_{s,\hat\mu}{\rm
Tr}\Big(\sigma_3 U(s,\mu) \sigma_3 U^{\dagger}(s,\mu)\Big)\,. \label{R}
\eeqn
($\sigma_3$ is the Pauli matrix). In the continuum limit a local condition of
the maximization can be written in the form of the differential equation,
$(\partial_{\mu}+igA_{\mu}^3)(A_{\mu}^1-iA_{\mu}^2)=0$.

\section{Numerical results}
\label{sec:numerics}

We simulate the $SU(2)$ gauge fields on the lattices $L_s^3 \times 4$,
$L_s=12,14,16,24$ with $C$--periodic boundary conditions in space
directions~\cite{Wie92}. The $C$--periodicity corresponds to the
anti-periodicity for the Abelian gauge fields which is required by the Gauss
law\footnote{One can not introduce the creation operator of the charged
particle in a finite volume with periodic boundary
conditions.}~\cite{MonopoleCondensation1}. In the case of $SU(2)$ gauge group
the $C$--periodic boundary conditions mean that on the space boundary we have
$U_{x,\mu} \to \Omega^+ U_{x,\mu} \Omega,\,\,\Omega = i \sigma_2$.

We fix AP and $F_{12}$ gauges as it was described in the previous Section. The
results for the MA gauge (quoted below) are taken from
Ref.~\cite{MonopoleCondensation1}. To get the effective potential in the AP and
$F_{12}$ gauges we use 400 independent configurations of $SU(2)$ gauge fields
for each value of the gauge coupling $\beta$ at a given lattice volume. On each
configuration the distribution of the monopole creation operator is evaluated
in 20 points. The logarithm of the distribution provides us with the effective
potential~\eq{eq:V}.

To evaluate the numerical errors of the potential we use the bootstrap method.
For each value of $\beta$ and the lattice volume we get a distribution of the
values of the monopole creation operator ("initial ensemble"). Then we use the
initial ensemble to construct a number (typically $N_{ens} = 500$) of
additional ensembles of the operator values by randomly choosing the entries
from the initial ensemble. Each value from the initial ensemble may enter the
additional ensemble more than one time. The number of entries in each of the
additional ensembles is the same as in the initial one. Then for each ensemble
we construct a histogram, the (minus) logarithm of which has a meaning of the
monopole potential, $V(\Phi)$, according to Eq.~\eq{eq:V}. Thus, for each value
of the monopole field, $\Phi$, we get $N_{ens}$ values of the potential, $V$,
which form a Gaussian distribution. The central value of this distribution
gives us the value of the potential at given lattice volume $L_s^2\times L_t$,
$\beta$ and $\Phi$, $V = V(\Phi)$, while the width of the distribution is the
statistical error.

\begin{figure}[!htb]
\begin{center}
\vspace{5mm}
\begin{tabular}{cc}
\includegraphics[angle=-00,height=6cm,width=78mm,clip=true]{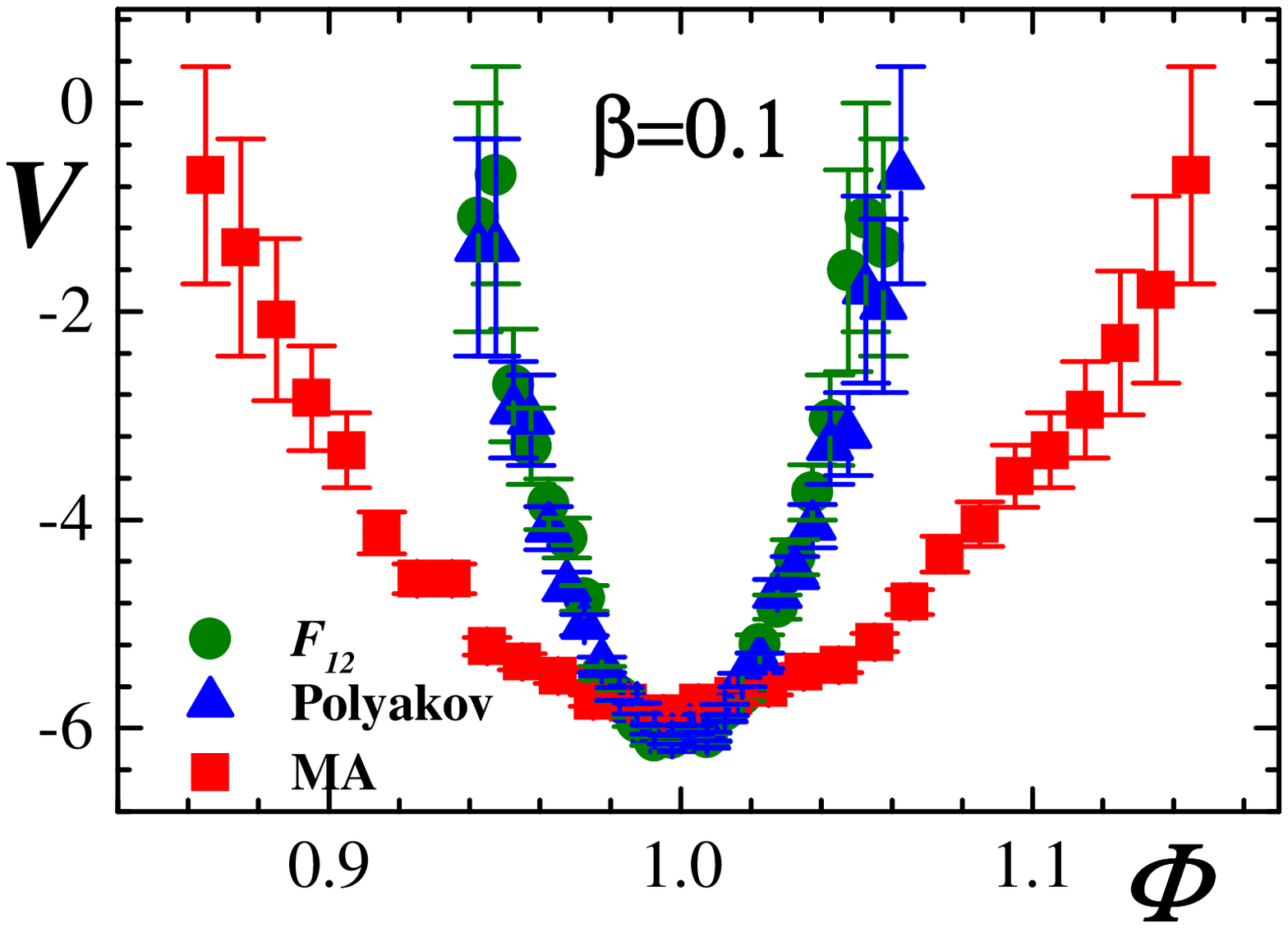} &
\includegraphics[angle=-00,height=6cm,width=78mm,clip=true]{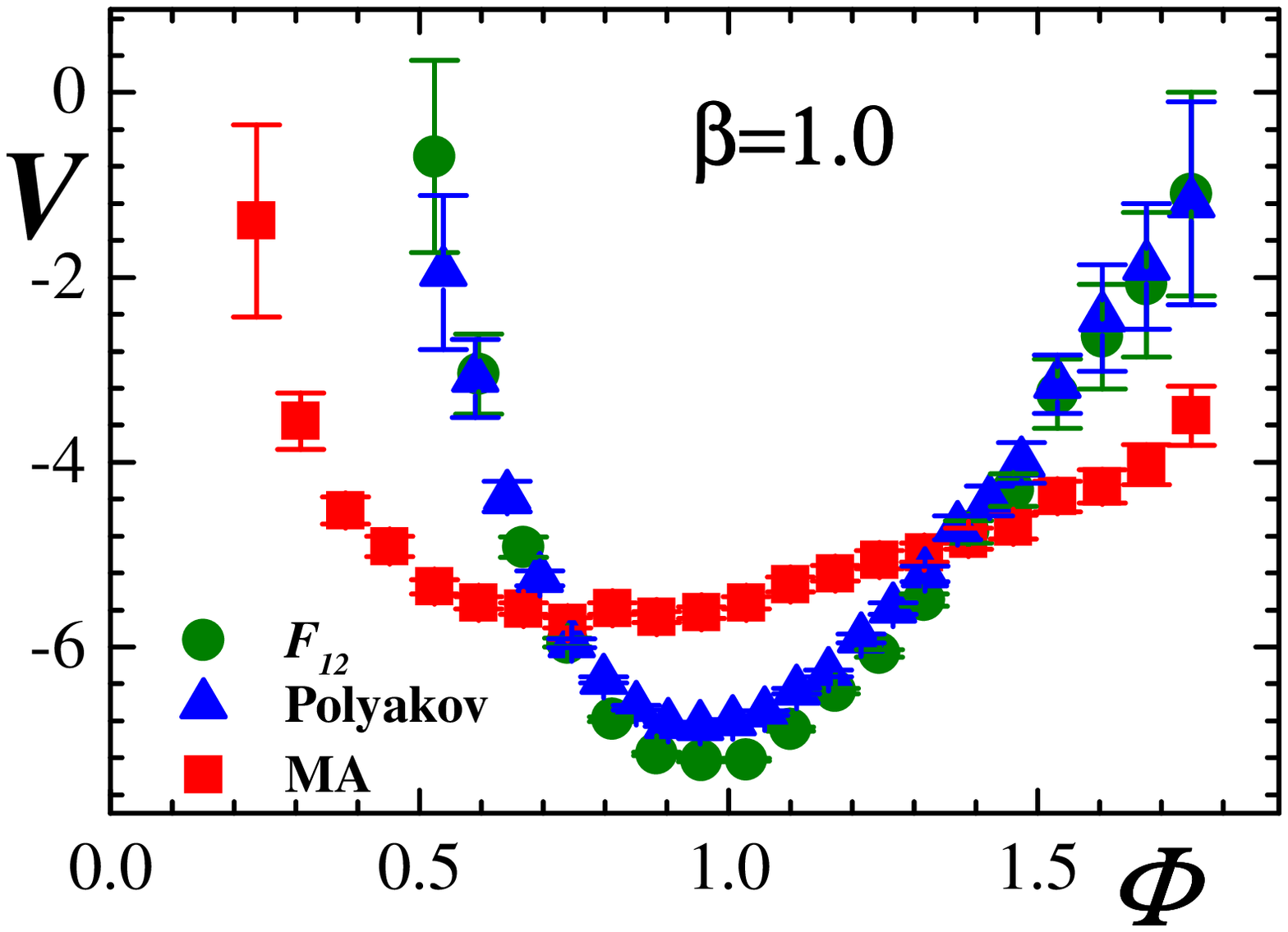} \\
(a) & (b) \vspace{5mm} \\
\includegraphics[angle=-00,height=6cm,width=78mm,clip=true]{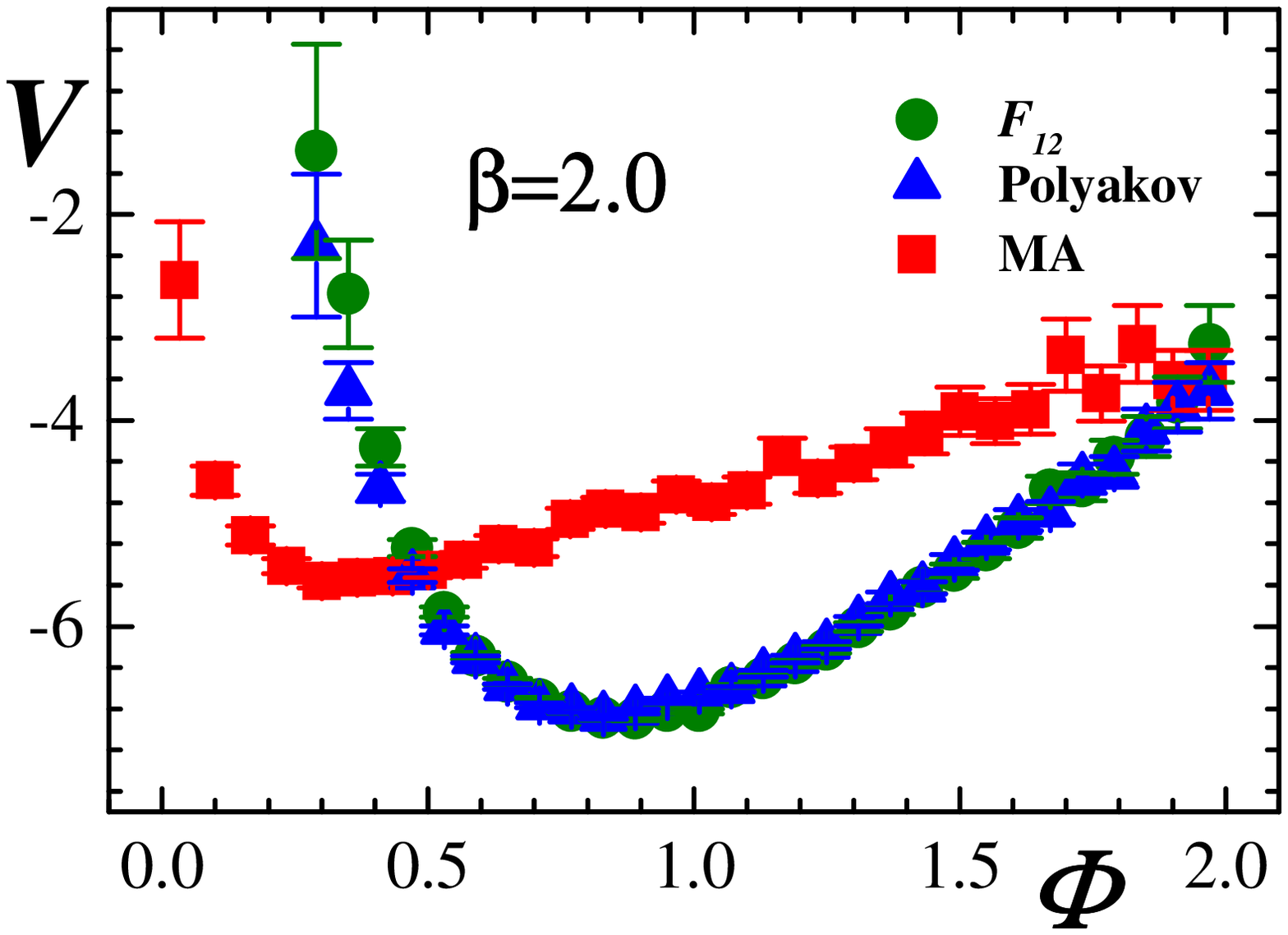} &
\includegraphics[angle=-00,height=6cm,width=78mm,clip=true]{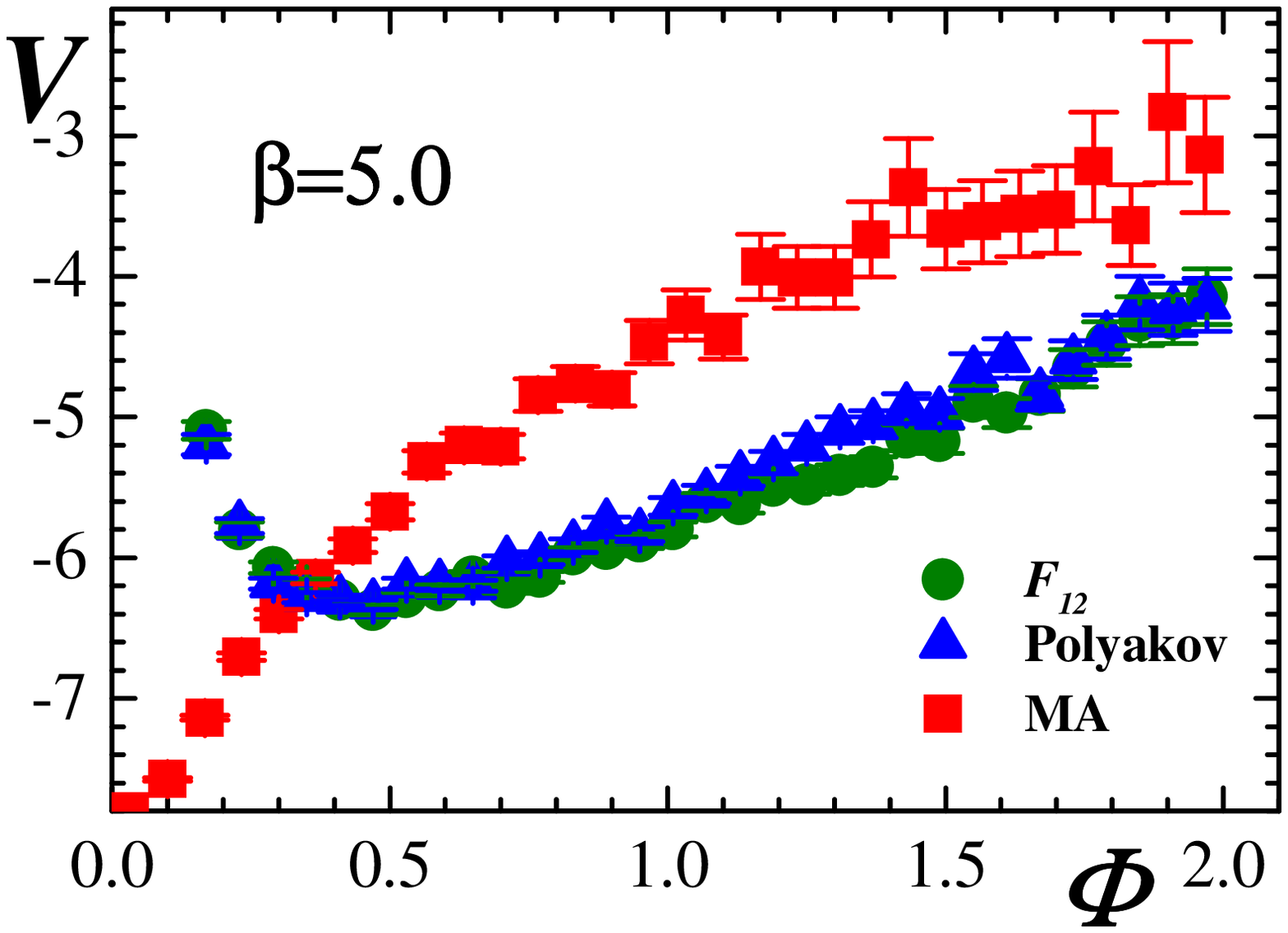} \\
(c) & (d)
\end{tabular}
\end{center}
\caption{The potential on the monopole field in the MA, AP and $F_{12}$
gauges at various values of the gauge coupling $\beta$. The data for the
MA gauge are taken from Ref.~\cite{MonopoleCondensation1}.}
\label{fig:examples}
\end{figure}
The examples of the effective monopole potentials for $16^3\times 4$ lattice
are shown on Figs.~\ref{fig:examples}(a-d). We depict the positive part of the
potentials ($\Phi > 0$) at various values of the gauge coupling $\beta$ in the
MA, AP and $F_{12}$ gauges. The minimum of the effective potential corresponds
to the value of the monopole condensate. The critical gauge coupling
corresponding to the temperature phase transition at our lattices is $\beta_c
\approx 2.3$. Thus Figs.~\ref{fig:examples}(a,b,c) correspond to the
confinement phase while Fig.~\ref{fig:examples}(d) corresponds to the
deconfinement phase.

First of all we note that for all considered values of $\beta$ (i) the minima
of the potentials in the AP and $F_{12}$ gauges coincide with each other within
numerical errors; (ii) the potential in the MA gauge is different from AP and
$F_{12}$ potentials. According to Fig.~\ref{fig:examples}(a) in the strong
coupling limit ($\beta = 0.1$) the minima of the monopole potential in all
three gauges are located at the same point, $\Phi_{\min} \approx 1$. As we
increase $\beta$ the difference in the monopole condensates in MA gauge and in
AP and $F_{12}$ gauges appears, see Fig.~\ref{fig:examples}(b,c). Moreover, in
the deep deconfinement phase, Fig.~\ref{fig:examples}(d), the monopole
condensate vanishes in MA gauge while in AP and $F_{12}$ gauges the condensate
is non--zero.

\begin{figure}[!htb]
\begin{center}
\vspace{5mm}
\begin{tabular}{cc}
\includegraphics[angle=-00,height=6cm,width=78mm,clip=true]{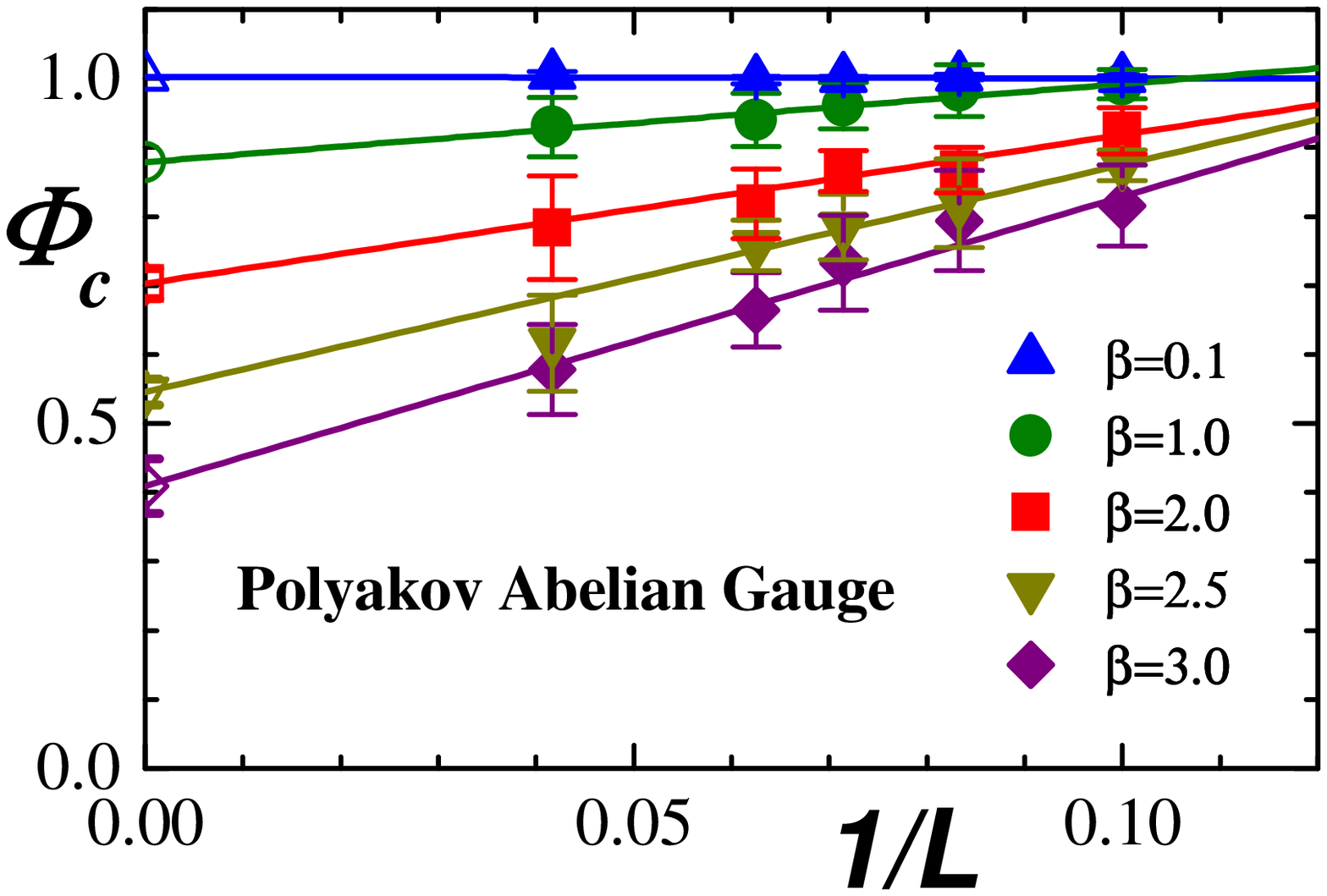} &
\includegraphics[angle=-00,height=6cm,width=78mm,clip=true]{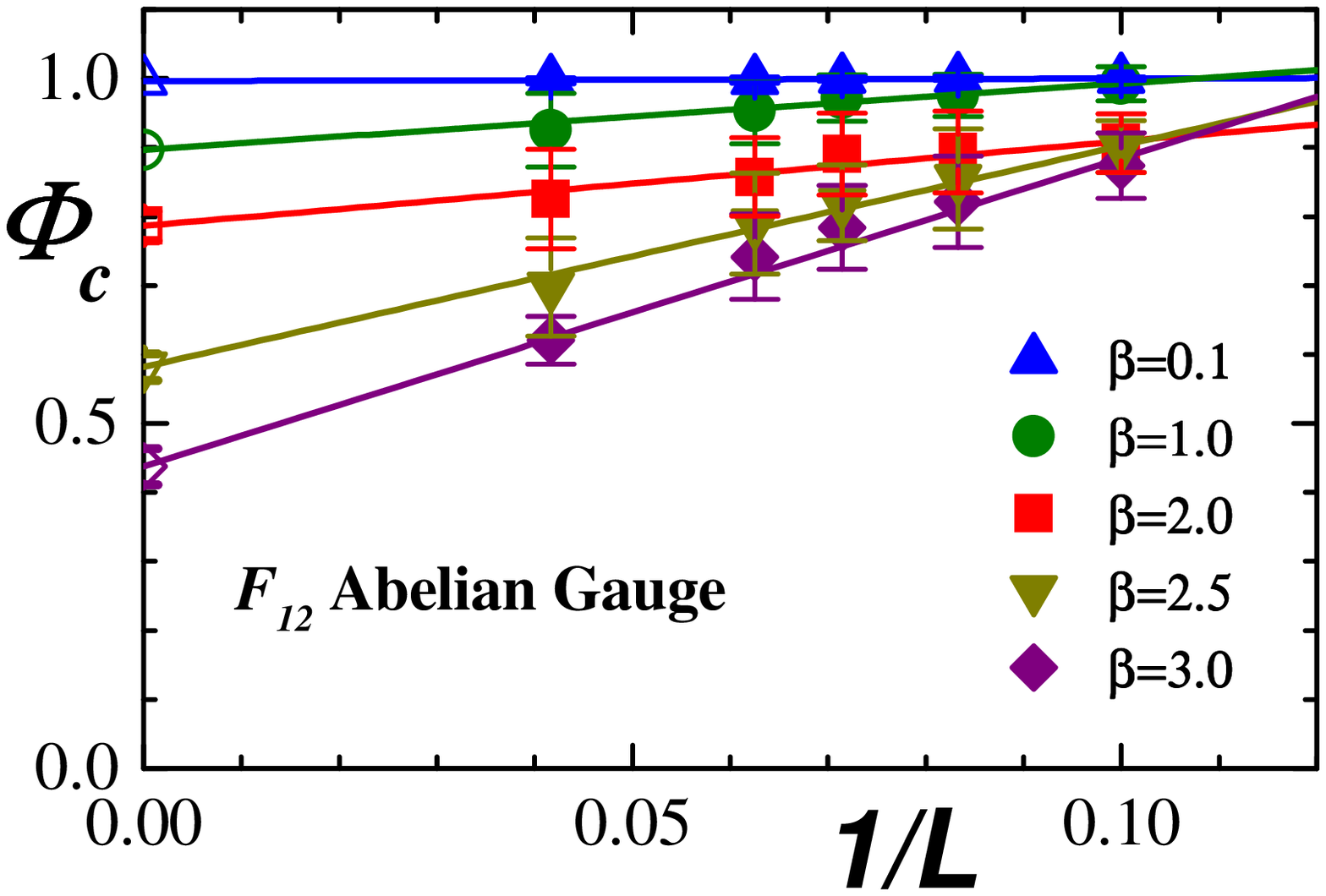} \\
(a) & (b)
\end{tabular}
\end{center}
\caption{The extrapolation~\eq{eq:extr} of the monopole condensate to
the thermodynamic limit in the AP and the $F_{12}$ gauges for various
values of the gauge coupling $\beta$. The spatial extensions of the lattices
are $L_s=12,14,16,24$.}
\label{fig:extr}
\end{figure}

Since the phase transition in  the $SU(2)$ gauge theory is of the second order,
the finite volume effects may be essential for the determination of the
monopole condensate. To get rid of finite volume corrections we measure the
condensate on the lattices with various spatial extensions ($L_s=12,14,16,24$)
and extrapolate the value of the condensate to the thermodynamic limit, $L_s
\to \infty$:
\beqn
\Phi_c = \Phi^{\mathrm{inf}}_c + \frac{C}{L}\,. \label{eq:extr}
\eeqn
The examples of the fits for the AP and the $F_{12}$ gauges are shown in
Figs.~\ref{fig:extr}(a,b). The values of $\chi^2/d.o.f.$ are in the range $0.2
\sim 1$.

\begin{figure}[!htb]
\begin{center}
\vspace{10mm}
\includegraphics[angle=-00,scale=0.6,clip=true]{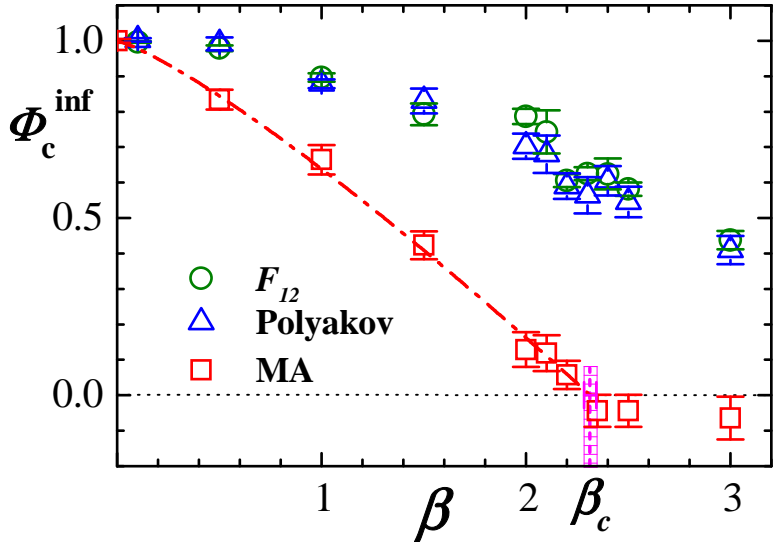}
\end{center}
\caption{The monopole condensate in the thermodynamic limit in
the MA, the AP and the $F_{12}$ gauges. The dash--dotted line is the fit
of the monopole condensate in the MA gauge by Eq.~\eq{eq:fit}. The critical
value of the gauge coupling (along with the numerical error) is denoted by
the vertical dashed line.}
\label{fig:inf}
\end{figure}

The monopole condensates in the thermodynamic limit ($L_s \to \infty$) for all
three Abelian projections are shown in Fig.~\ref{fig:inf} as functions of
$\beta$. One can clearly see that the monopole condensate in the MA projection
vanishes at a certain critical $\beta = \beta_c$ which is very close to the
phase transition point, $\beta_c \approx 2.3$. Contrary to the MA gauge the
monopole condensates obtained in the AP and the $F_{12}$ gauges do not vanish
at $\beta = \beta_c$. This result is in contradiction with observations of
Ref.~\cite{DiGiacomo}.

The dependence of the monopole condensate on $\beta$ can be fitted by
the following function:
\beqn
\Phi^{\mathrm{inf}}_c (\beta) = 1 - (\beta/\beta_c)^\gamma\,,
\label{eq:fit}
\eeqn
with $\chi^2/d.o.f. \approx 0.3$. It occurs that $\gamma = 1.2(5)$ and $\beta_c
= 2.31(3)$. The value of $\beta_c$ coincides within error bars with the known
critical value~\cite{L3x4} on $L_s^3 \times 4$ lattice.

\clearpage

\section{Conclusion}

Summarizing, we have presented an evidence that the monopole condensate in
different Abelian projections coincide with each other only in the (unphysical)
strong coupling region. Generally, the condensate depends on the choice of the
Abelian projection. We have considered three Abelian projections and only in MA
projection the vacuum behaves as the dual superconductor. Our results are in
contradiction with conclusions of Ref.~\cite{DiGiacomo} where condensate was
found to be projection--independent.

\section*{Acknowledgments}
This work was supported by the grants
RFBR 02-02-17308,          
RFBR 01-02-17456,          
RFBR-DFG-03-02-04016,
DFG-RFBR 436 RUS 113/739/0,      
INTAS-00-00111,            
CRDF award RPI-2364-MO-02, 
and MK-4019.2004.2.        

\end{document}